# A Swarm Intelligence Based Scheme for Complete and Fault-tolerant Identification of a Dynamical Fractional Order Process

Deepyaman Maiti, Ayan Acharya, and Amit Konar
Department of Electronics and Telecommunication Engineering
Jadavpur University
Kolkata, India
deepyamanmaiti@gmail.com, masterayan@gmail.com, and konaramit@yahoo.co.in

*Abstract*—System identification refers to estimation of process parameters and is a necessity in control theory. Physical systems usually have varying parameters. For such processes, accurate identification is particularly important. Online identification schemes are also needed for designing adaptive controllers. Real processes are usually of fractional order as opposed to the ideal integral order models. In this paper, we propose a simple and elegant scheme of estimating the parameters for such a fractional order process. A population of process models is generated and updated by particle swarm optimization (PSO) technique, the fitness function being the sum of squared deviations from the actual set of observations. Results show that the proposed scheme offers a high degree of accuracy even when the observations are corrupted to a significant degree. Additional schemes to improve the accuracy still further are also proposed and analyzed.

I. INTRODUCTION

The first references on fractional order derivatives were made in the 17$^{th}$ century. Since then, the theory of fractional order integration and derivatives has been highly developed by many mathematicians. In the last five decades, many authors made a great effort to apply this knowledge in practice. But only in the last decade can we find some significant works concerned with the description, analysis, and synthesis of fractional-order regulated systems.

Proper estimation of the parameters of a real process, fractional or otherwise, is a challenge to be encountered in the context of system identification [1], [2]. Many statistical and geometric methods such as least square and regression models are widely used for real-time parameter estimation.

The problem of parameter estimation becomes more difficult for a fractional order system compared to an integral order one. The real world objects or processes that we want to estimate are generally of fractional order [3]. A typical example of a non-integer (fractional) order system is the voltage-current relation of a semi-infinite lossy RC line or diffusion of heat into a semi-infinite solid, where heat flow q(t) is equal to the half-derivative of temperature T(t).

So far, however, the usual practice when dealing with a fractional order process has been to use an integer order approximation. Disregarding the fractional order of the system was caused mainly by the non-existence of simple mathematical tools for the description of such systems. Since major advances have been made in this area recently, it is possible to consider also the real order of the dynamical systems. Such models are more adequate for the description of dynamical systems with distributed parameters than integer-order models with concentrated parameters. Most classical identification methods cannot cope with fractional order transfer functions. Yet, this challenge must be overcome if we want to design a proper adaptive or self-tuning fractional order controller. Need for design of adaptive controllers gives an impetus to finding accurate schemes for system identification.

Computation of transfer characteristics of the fractional order dynamic systems has been the subject of several publications [4] – [6], e.g. by numerical methods [4], as well as by analytical methods [5]. In this paper, we propose a general method for the estimation of parameters of a fractional order system using PSO technique. PSO, a stochastic optimization strategy from the family of evolutionary computation, is a biologically inspired technique originally proposed in [7]. We use PSO to find the process model whose outputs match the set of observations from the actual fractional order system most closely. This method enables us to work with the actual fractional order process rather than an integer order approximation. Using it in a system with known parameters will do the verification of the correctness of the identification.

Although the direct application of the PSO algorithm gives very accurate estimations, we will propose algorithms to check the accuracy of these results and improve on them. The theory of fractional calculus is needed in order to realize the significance of a fractional order system, which must consist of fractional order integrators and differentiators.

## II. FRACTIONAL CALCULUS THEORY AND PARTICLE SWARM OPTIMIZATION TECHNIQUE

### A. Theory of Fractional Calculus

The fractional calculus is a generalization of integration and derivation to non-integer order operators. At first, we generalize the differential and integral operators into one fundamental operator ${}_aD_t^\alpha$ where:

$$ {}_aD_t^\alpha = \frac{d^\alpha}{dt^\alpha} \text{ for } \Re(\alpha) > 0; = 1 \text{ for } \Re(\alpha) = 0; = \int_a^t (d\tau)^{-\alpha} $$

for $\Re(\alpha) > 0$. (1)

The two definitions used for fractional differintegral are the Riemann-Liouville definition and the Grunwald-Letnikov definition. The Grunwald-Letnikov definition is

$$ {}_aD_t^\alpha f(t) = \lim_{h \to 0} \frac{1}{h^\alpha} \sum_{j=0}^{\left[\frac{t-a}{h}\right]} (-1)^j \binom{\alpha}{j} f(t - jh) \quad (2) $$

where $[y]$ means the greatest integer not exceeding y.

Derived from the Grunwald-Letnikov definition, the numerical calculation formula of fractional derivative can be achieved as:

$$ {}_{t-L}D_t^\alpha x(t) \approx h^{-\alpha} \sum_{j=0}^{[L/T]} b_j x(t - jh) \quad (3) $$

where L is the length of memory. T, the sampling time always replaces the time increment h during approximation. The weighting coefficients $b_j$ can be calculated recursively by:

$$ b_0 = 1, b_j = \left(1 - \frac{1+\alpha}{j}\right) b_{j-1}, (j \geq 1). \quad (4) $$

### B. Particle Swarm Optimization Technique

The PSO algorithm [7] - [9] attempts to mimic the natural process of group communication of individual knowledge, which occurs when a social swarm elements flock, migrate, forage, etc. in order to achieve some optimum property such as configuration or location.

The 'swarm' is initialized with a population of random solutions. Each particle in the swarm is a different possible set of the unknown parameters to be optimized. Representing a point in the solution space, each particle adjusts its flying toward a potential area according to its own flying experience and shares social information among particles. The goal is to efficiently search the solution space by swarming the particles toward the best fitting solution encountered in previous iterations with the intent of encountering better solutions through the course of the process and eventually converging on a single minimum error solution.

Let the swarm consist of N particles moving around in a D-dimensional search space. Each particle is initialized with a random position and a random velocity. Each particle modifies its flying based on its own and companions' experience at every iteration. The $i^{th}$ particle is denoted by $X_i$, where $X_i = (x_{i1}, x_{i2}, \ldots, x_{iD})$. Its best previous solution (pbest) is represented as $P_i = (p_{i1}, p_{i2}, \ldots, p_{iD})$. Current velocity (position changing rate) is described by $V_i$, where $V_i = (v_{i1}, v_{i2}, \ldots, v_{iD})$. Finally, the best solution achieved so far by the whole swarm (gbest) is represented as $P_g = (p_{g1}, p_{g2}, \ldots, p_{gD})$.

At each time step, each particle moves towards pbest and gbest locations. The fitness function evaluates the performance of particles to determine whether the best fitting solution is achieved. The particles are manipulated according to the following equations:

$$ v_{id}(t+1) = \omega v_{id}(t) + c_1 \varphi_1 (p_{id}(t) - x_{id}(t)) + c_2 \varphi_2 (p_{gd}(t) - x_{id}(t)) \quad (5) $$

$$ x_{id}(t+1) = x_{id}(t) + v_{id}(t+1). \quad (6) $$

(The equations are presented for the $d^{th}$ dimension of the position and velocity of the $i^{th}$ particle.)

Here, $c_1$ and $c_2$ are two positive constants, called cognitive learning rate and social learning rate respectively, $\varphi_1$ and $\varphi_2$ are two random functions in the range [0,1], $\omega$ is the time-decreasing inertia factor designed by Eberhart and Shi [8]. The inertia factor balances the global wide-range exploitation and the nearby exploration abilities of the swarm.

## III. APPLICATION OF THE PSO ALGORITHM TO THE PROBLEM OF PARAMETER IDENTIFICATION

We have considered a fractional process whose transfer function is of the form $\frac{1}{a_1 s^\alpha + a_2 s^\beta + a_3}$. It should be noted that without loss of generality, we may presume the dc gain to be unity so that the dc gain and its possible fluctuations are included in the coefficients $a_1$, $a_2$ and $a_3$. This system has five varying parameters, namely three coefficients $a_1$, $a_2$ and $a_3$, and two fractional powers $\alpha$ and $\beta$.

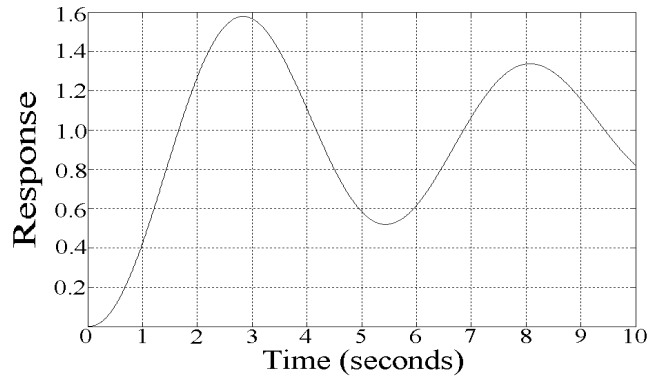

Figure 1. Unit step response of the system to be identified

We apply R(s)=1/s (unit step) to the actual system and obtain sampled values of output c(t). The PSO algorithm will search the solution space to come up with a process model which replicates the observed c(t) values for the same input signal, i.e. unit step. Let, for a process model, the output response for unit step input (obtained by numerical inverse laplace operation) is p(t). We will define a parameter $F = \sum_t [c(t) - p(t)]^2$, which gives a measure of the deviation of the output of the trial process model from the output of the actual process. F is the fitness function that the PSO algorithm will try to minimize. At F = 0, the unknown parameters are optimized. The position vector of the best particle, i.e. the optimized value of $\{a_1, \alpha, a_2, \beta, a_3\}$ is the identified parameter set. The process model corresponding to the optimized solution set should provide output identical to c(t) for unit step input.

Clearly, our only source of information about the actual process is the set of output readings from it. The PSO algorithm will try to find a process model that matches these readings. So we have to perform one transformation from s-domain to discrete time domain, since the actual readings will obviously be in time domain. An alternative approach is to convert all data into z-domain.

There are five unknown parameters to be optimized. So the present problem of system identification can be solved by a straightforward application of the PSO algorithm for optimization on a five-dimensional solution space. This approach gives very accurate results, even when the input data is corrupted to a significant extent. We will first study this scheme under both ideal and non-ideal (random error component added to readings) conditions, and then propose and analyze suitable methods for further improvements.

We will start our analysis by first assuming that only four of the five parameters are varying while the other is constant and known beforehand. Then we will increase the number of varying parameters to five. As we increase the number of varying parameters, we will observe the accuracy of identification, the number of PSO particles and the number of runs needed for convergence, and also the position and velocity limits used. Section IIIA studies the identification process when only four of the parameters are varying. Section IIIB studies the identification process when all five parameters are varying. In each section we have considered both ideal and erroneous scenarios. The PSO parameters used are: the inertia factor ω decreases linearly from 0.9 to 0.4, the learning rates $c_1$=1.4, $c_2$=1.4. These are kept constant for all the runs of the PSO algorithm. The other parameters are mentioned at suitable places.

*A. Identification for Four Varying Parameters: $a_1$, α, $a_2$ and $a_3$*

The process whose parameters are to be estimated is $\frac{1}{a_1 s^\alpha + a_2 s^{0.9} + a_3}$, i.e. β = 0.9 is known beforehand. Synthetic data for c(t) [input: unit step] is created using $a_1$ = 0.8, α = 2.2, $a_2$ = 0.5, $a_3$ = 1. That is, the values of c(t) at different time instants are obtained assuming a process with transfer function $\frac{1}{0.8 s^{2.2} + 0.5 s^{0.9} + 1}$.

The solution space is four-dimensional. So each particle of the population has a four-dimensional position vector $(a_1, \alpha, a_2, a_3)$. The velocity vectors, the personal bests and the global best are also four-dimensional.

Sampling frequency is 20 samples per second. We will consider two cases: first when the readings c(t) are accurate, and second, when the readings are erroneous (corrupted). Error in taking readings of c(t) is simulated by adding a random number in the range [−0.05, 0.05] to each reading. This error component is quite considerable since the magnitude of the output response is usually below unity. We will show that the estimations of parameters are quite accurate in spite of the erroneous readings and the low sampling rate.

Number of particles in the population is 40. The PSO algorithm is run for 150 iterations, and this is kept as the stop criterion. The search ranges are: $a_1$, $a_2$, $a_3$: 0 to 2.0 (for all three parameters) and α: 2.0 to 2.4. The velocity ranges are: $a_1, a_2, a_3$: –0.5 to 0.5 (for all three parameters) and α: -0.1 to 0.1.

*1) Identification when Random Error Component is NOT Added (Ideal Case)*

Tables I and II give the results for five *independent and consecutive* runs of the PSO algorithm when the readings are accurate (i.e. random corruption component is not added). Results are shown up to four places after the decimal.

TABLE I. ESTIMATED PARAMETER SETS AND FITNESS

| Estimated Parameters | | | | Fitness F |
|---|---|---|---|---|
| $a_1$ | α | $a_2$ | $a_3$ | |
| 0.8000 | 2.2000 | 0.5000 | 1.0000 | 0 |
| 0.8000 | 2.2000 | 0.5001 | 1.0000 | 1.6381 x $10^{-7}$ |
| 0.8000 | 2.2000 | 0.5000 | 1.0000 | 0 |
| 0.8000 | 2.2000 | 0.5000 | 1.0000 | 0 |
| 0.8000 | 2.2000 | 0.5000 | 1.0000 | 0 |

TABLE II. STATISTICAL PROPERTIES OF ESTIMATED PARAMETER SETS

| | $a_1$ | α | $a_2$ | $a_3$ |
|---|---|---|---|---|
| Mean of 5 Runs | 0.8000 | 2.2000 | 0.5000 | 1.0000 |
| Std. Deviation | 0 | 0 | 4.4721 x $10^{-5}$ | 0 |
| Pc Error of Mean | 0 | 0 | 0.0040 | 0 |

We note that the best estimated parameter set (corresponding to the least value of F) is $a_1 = 0.8000$, $\alpha = 2.2000$, $a_2 = 0.5000$, $a_3 = 1.0000$. The percentage errors are respectively 0, 0, 0 and 0. The best (least) value of F is F = 0.

*2) Identification when Random Error Component is Added*

Tables III and IV give the results for five *independent and consecutive* runs of the PSO algorithm when a random corruption component in the range [$-0.05$, $0.05$] is added to each reading of c(t). So in each run the PSO algorithm tries to identify a process model that is deviated from the actual process, the deviation randomly varying from run to run.

TABLE III.  ESTIMATED PARAMETER SETS AND FITNESS

| Estimated Parameters | | | | Fitness F |
|---|---|---|---|---|
| $a_1$ | $\alpha$ | $a_2$ | $a_3$ | |
| 0.8015 | 2.1999 | 0.5019 | 1.0006 | 0.1704 |
| 0.7950 | 2.1865 | 0.4803 | 0.9991 | 0.1620 |
| 0.8050 | 2.2291 | 0.5466 | 0.9932 | 0.1707 |
| 0.7999 | 2.2046 | 0.5000 | 0.9991 | 0.1705 |
| 0.8000 | 2.1967 | 0.5023 | 0.9988 | 0.1543 |

TABLE IV.  STATISTICAL PROPERTIES OF ESTIMATED PARAMETER SETS

| | $a_1$ | $\alpha$ | $a_2$ | $a_3$ |
|---|---|---|---|---|
| Mean of 5 Runs | 0.8003 | 2.2034 | 0.5062 | 0.9982 |
| Std. Dev. | 0.0036 | 0.0158 | 0.0244 | 0.0029 |
| Pc Error | 0.0375 | 0.1545 | 1.2400 | 0.1900 |

We note that the best estimated parameter set (corresponding to the least value of F) is $a_1 = 0.8000$, $\alpha = 2.1967$, $a_2 = 0.5023$, $a_3 = 0.9988$. The percentage errors are respectively 0.0061, 0.1492, 0.4518 and 0.1186. The best (least) value of F is F = 0.1543.

*B.  Identification for Five Varying Parameters: $a_1$, $\alpha$, $a_2$, $\beta$ and $a_3$*

The process whose parameters are to be estimated is $\frac{1}{a_1 s^\alpha + a_2 s^\beta + a_3}$. Synthetic data for c(t) [input: unit step] is created using $a_1$ = 0.8, $\alpha$ = 2.2, $a_2$ = 0.5, $\beta$ = 0.9, $a_3$ = 1. Sampling frequency is 20 samples per second. Error in taking readings of c(t) is simulated by adding a random number in the range [$-0.05$, $0.05$] to each reading.

Number of particles in the population is 50. The PSO algorithm is run for 200 iterations, and this is kept as the stop criterion. The search ranges are: $a_1$, $a_2$, $a_3$: 0 to 2.0 (for all three parameters), α: 2.0 to 2.4 and β: 0.7 to 1.1. The velocity ranges are: $a_1$, $a_2$, $a_3$: –0.5 to 0.5 (for all three parameters), α, β: -0.1 to 0.1 (for both parameters).

*1) Identification when Random Error Component is NOT Added (Ideal Case)*

Tables V and VI give the results for five *independent and consecutive* runs of the PSO algorithm when the readings are accurate (i.e. random corruption component is not added). Results are shown up to four places after the decimal.

TABLE V.  ESTIMATED PARAMETER SETS AND FITNESS

| Estimated Parameters | | | | | Fitness F |
|---|---|---|---|---|---|
| $a_1$ | $\alpha$ | $a_2$ | $\beta$ | $a_3$ | |
| 0.7989 | 2.2015 | 0.5014 | 0.9029 | 1.0004 | 7.3110 x $10^{-6}$ |
| 0.8066 | 2.1892 | 0.4897 | 0.8802 | 0.9973 | 4.2457 x $10^{-4}$ |
| 0.8047 | 2.1955 | 0.4969 | 0.8878 | 0.9981 | 1.3228 x $10^{-4}$ |
| 0.8015 | 2.1983 | 0.4987 | 0.8961 | 0.9995 | 1.0938 x $10^{-5}$ |
| 0.7474 | 2.2558 | 0.5452 | 1.0175 | 1.0156 | 0.0110 |

TABLE VI.  STATISTICAL PROPERTIES OF ESTIMATED PARAMETER SETS

| | $a_1$ | $\alpha$ | $a_2$ | $\beta$ | $a_3$ |
|---|---|---|---|---|---|
| Mean | 0.7918 | 2.2081 | 0.5064 | 0.9169 | 1.0022 |
| Std. Dev. | 0.0250 | 0.0271 | 0.0221 | 0.0569 | 0.0076 |
| Pc. Error | 1.0250 | 0.3682 | 1.2800 | 1.8778 | 0.2200 |

We note that the best estimated parameter set (corresponding to the least value of F) is $a_1 = 0.7989$, $\alpha = 2.2015$, $a_2 = 0.5014$, $\beta = 0.9029$, $a_3 = 1.0004$. The percentage errors are respectively 0.1390, 0.0697, 0.2717, 0.3266 and 0.0385. The best (least) value of F is 7.3110 x $10^{-6}$.

*2) Identification when Random Error Component is Added*

Tables VII and VIII give the results for five *independent and consecutive* runs of the PSO algorithm when a random corruption component in the range [$-0.05$, $0.05$] is added to each reading of c(t).

TABLE VII.  ESTIMATED PARAMETER SETS AND FITNESS

| Estimated Parameters | | | | | Fitness F |
|---|---|---|---|---|---|
| $a_1$ | $\alpha$ | $a_2$ | $\beta$ | $a_3$ | |
| 0.7938 | 2.2061 | 0.5033 | 0.9249 | 1.0045 | 0.1533 |
| 0.7882 | 2.2219 | 0.5206 | 0.9296 | 1.0004 | 0.1645 |
| 0.8105 | 2.1842 | 0.4869 | 0.8600 | 0.9938 | 0.1714 |
| 0.7789 | 2.2091 | 0.4966 | 0.9514 | 1.0098 | 0.1599 |
| 0.7632 | 2.2373 | 0.5314 | 0.9828 | 1.0121 | 0.1589 |

TABLE VIII.  STATISTICAL PROPERTIES OF ESTIMATED PARAMETER SETS

| | $a_1$ | $\alpha$ | $a_2$ | $\beta$ | $a_3$ |
|---|---|---|---|---|---|
| Mean | 0.7869 | 2.2117 | 0.5078 | 0.9297 | 1.0041 |
| Std. Dev. | 0.0175 | 0.0197 | 0.0181 | 0.0452 | 0.0074 |
| Pc. Error | 1.6375 | 0.5318 | 1.5600 | 3.3000 | 0.4100 |

We note that the best estimated parameter set (corresponding to the least value of F) is $a_1 = 0.7938$, $\alpha = 2.2061$, $a_2 = 0.5033$, $\beta = 0.9249$, $a_3 = 1.0045$. The percentage errors are respectively 0.7768, 0.2764, 0.6698, 2.7646 and 0.4498. The best (least) value of F is F = 0.1533.

### C. Analysis of the Scheme of Direct Application of the PSO Algorithm for Parameter Estimation

This scheme offers highly accurate results when the number of varying parameters is limited to three or four, i.e. when at least one or two parameters are known to remain stationary. The accuracy then routinely touches 100 percent. The results are very promising even when significant amounts of random error are added. However when all the parameters are varying, this method does not guarantee cent percent accuracy. So in sections IV and V, we will propose and analyze a scheme to check the results of a straightforward application of a five-parameter optimization PSO algorithm, and also a method to circumvent the inaccuracy in estimates that arises due to all the system parameters being of a dynamic nature.

### IV. VERIFICATION OF THE RESULTS OF A DIRECT APPLICATION OF A FIVE-PARAMETER OPTIMIZATION PSO ALGORITHM USING FRACTIONAL CALCULUS TECHNIQUES

We will assume that the fractional powers have been correctly identified by the PSO algorithm, and using those values of the fractional powers, we will check the coefficient terms i.e. $a_1$, $a_2$, $a_3$ and the resulting fitness value by means of fractional calculus theory. We have considered a fractional process whose transfer function is of the form $\frac{1}{a_1 s^\alpha + a_2 s^\beta + a_3}$. If C(s) be the output and R(s) the input, we will have: $\frac{C(s)}{R(s)} = \frac{1}{a_1 s^\alpha + a_2 s^\beta + a_3}$,

$\Rightarrow R(s) = a_1 s^\alpha C(s) + a_2 s^\beta C(s) + a_3 C(s)$.

In time domain, $r(t) = a_1 D^\alpha c(t) + a_2 D^\beta c(t) + a_3 c(t)$ (7)

$$\Rightarrow r(t) \approx a_1 T^{-\alpha} \sum_{j=0}^{[L/T]} b_j c(t-jT) + a_2 T^{-\beta} \sum_{j=0}^{[L/T]} b_j c(t-jT) + a_3 c(t)$$ (8)

The proposed scheme requires sampled input at time instant t and sampled outputs at time instants $t$, $t-T$, $t-2T$, $t-3T$, ……. Sampled outputs are required for a time length L previous to t, T being the sampling time. Calculation of fractional derivatives and integrals requires the past history of the process to be remembered. So value of "L" should be high.

Values of $D^\alpha c(t)$ and $D^\beta c(t)$ can thus be calculated so that (7) reduces to the form $a_1 p + a_2 q + a_3 r = s$, where $p, q, r, s$ are constants whose values have been determined.

Let us assume that we have a set of sampled outputs c(t) from the system for unit step test signal. That is, we have
$u(t) = a_1 D^\alpha c(t) + a_2 D^\beta c(t) + a_3 c(t)$. (9)

Now there are three unknown parameters, namely $a_1$, $a_2$ and $a_3$. So we need three simultaneous equations to solve for them. One equation is (9). We will integrate both sides of (9) to get $\int u(t)dt = \int [a_1 D^\alpha c(t) + a_2 D^\beta c(t) + a_3 c(t)]dt$ which gives us $r(t) = a_1 D^{\alpha-1} c(t) + a_2 D^{\beta-1} c(t) + a_3 D^{-1} c(t)$ (10) where r(t) signifies unit ramp input and c(t) is the output due to unit step input. Thus we have derived a second equation relating $a_1$, $a_2$ and $a_3$.

The third equation will be obtained by integrating both sides of (10). This gives us
$p(t) = a_1 D^{\alpha-2} c(t) + a_2 D^{\beta-2} c(t) + a_3 D^{-2} c(t)$ (11)
where p(t) signifies parabolic input and c(t) is the output due to unit step input.

It can be seen that (9), (10), (11) are three distinct equations in $a_1$, $a_2$ and $a_3$. So we can solve them simultaneously to identify the three unknown parameters $a_1$, $a_2$ and $a_3$. Now let us suppose we wish to verify the following estimates (from runs 1, 3, 5 as given in table V).

TABLE IX. ESTIMATED PARAMETERS AND FITNESS FROM RUNS 1, 3 AND 5 OF TABLE V

| Estimated Parameters | | | | | Fitness F |
|---|---|---|---|---|---|
| $a_1$ | $\alpha$ | $a_2$ | $\beta$ | $a_3$ | |
| 0.7989 | 2.2015 | 0.5014 | 0.9029 | 1.0004 | 7.3110 x 10$^{-6}$ |
| 0.8047 | 2.1955 | 0.4969 | 0.8878 | 0.9981 | 1.3228 x 10$^{-4}$ |
| 0.7474 | 2.2558 | 0.5452 | 1.0175 | 1.0156 | 0.0110 |

Using the method explained, we calculate the values of $a_1$, $a_2$ and $a_3$ and hence find the resulting fitnesses. Length of memory L = 10 seconds and T = 0.001 seconds is used to calculate the fractional derivatives. Table X shows the results.

TABLE X. ESTIMATED PARAMETER SETS AND FITNESS USING NUMERICAL FRACTIONAL DIFFERENTIATION

| Sl. No. | $\alpha$ | $\beta$ | $a_1$ | $a_2$ | $a_3$ | Fitness F |
|---|---|---|---|---|---|---|
| 1 | 2.2015 | 0.9029 | 0.8021 | 0.4994 | 1.0006 | 0.2146 |
| 2 | 2.1955 | 0.8878 | 0.8074 | 0.4928 | 0.9987 | 0.3066 |
| 3 | 2.2558 | 1.0175 | 0.7546 | 0.5603 | 1.0136 | 2.7921 |

From the fitness column of table X, we can check that the first model is the best identification, while the third is the worst. This result tallies with the result obtained by direct application of PSO (table IX).

## V. CONCENTRATED SEARCH ALGORITHM TO IMPROVE ACCURACY

The inaccuracies obtained in the results of section IIIB arise due to all the parameters being of a dynamic nature. In the concentrated search algorithm, we will select the parameter that has the lowest range of variation. We will subdivide this range of variation into a few subintervals, and assume as the nominal values the central points of those subintervals. Then we can employ the PSO algorithm to locate the subinterval in which the parameter lies by checking the fitness values. That subinterval will be divided into sub-subintervals and the PSO algorithm is run using the new nominal values. In this way we can obtain far more accurate estimates than was obtained by direct application of a five-parameter optimization scheme.

### A. Illustration of the Concentrated Search Algorithm

The range of variation of $\beta$ is 0.7 to 1.1. Let us subdivide this interval into four subintervals. The nominal values are 0.75, 0.85, 0.95 and 1.05. Now we optimize the remaining four parameters and note the fitness values in each of the four cases. The results are tabulated below.

TABLE XI.

| $\beta$ range | Nominal value of $\beta$ | Fitness F |
|---|---|---|
| 0.7 to 0.8 | 0.75 | 0.0145 |
| 0.8 to 0.9 | 0.85 | 0.0017 |
| 0.9 to 1.0 | 0.95 | 0.0026 |
| 1.0 to 1.1 | 1.05 | 0.0184 |

The best fitness corresponds to the subinterval 0.8 to 0.9. So we subdivide this subinterval and continue the process.

TABLE XII. THE BEST FITNESS CORRESPONDS TO INTERVAL 0.88 TO 0.90

| $\beta$ range | Nominal value of $\beta$ | Fitness F |
|---|---|---|
| 0.8 to 0.82 | 0.81 | 0.0054 |
| 0.82 to 0.84 | 0.83 | 0.0033 |
| 0.84 to 0.86 | 0.85 | 0.0017 |
| 0.86 to 0.88 | 0.87 | $6.3421 \times 10^{-4}$ |
| 0.88 to 0.90 | 0.89 | $7.8781 \times 10^{-5}$ |

TABLE XIII. THE BEST FITNESS CORRESPONDS TO INTERVAL 0.89 TO 0.90

| $\beta$ range | Nominal value of $\beta$ | Fitness F |
|---|---|---|
| 0.88 to 0.89 | 0.885 | $1.6045 \times 10^{-4}$ |
| 0.89 to 0.90 | 0.895 | $1.7971 \times 10^{-5}$ |

TABLE XIV.

| $\beta$ range | Nominal value of $\beta$ | Fitness F |
|---|---|---|
| 0.890 to 0.892 | 0.891 | $5.8041 \times 10^{-5}$ |
| 0.892 to 0.894 | 0.893 | $3.8822 \times 10^{-5}$ |
| 0.894 to 0.896 | 0.895 | $1.7971 \times 10^{-5}$ |
| 0.896 to 0.898 | 0.897 | $8.0263 \times 10^{-6}$ |
| 0.898 to 0.900 | 0.899 | $7.2117 \times 10^{-7}$ |

If we stop at this point, our best estimates of the parameters are $a_1 = 0.8004$, $\alpha = 2.1996$, $a_2 = 0.4997$, $\beta = 0.899$, $a_3 = 0.9999$. The percentage errors are respectively 0.0500, 0.0182, 0.0600, 0.1111 and 0.0100. The best (least) value of F is $F = 7.2117 \times 10^{-7}$. This result is much more accurate than the best result obtained from a direct application of the five parameter optimization PSO algorithm.

## VI. COMMENTS, COMPARISONS AND CONCLUSIONS

An elegant method for the estimation of the parameters of a fractional order system is proposed. The proposed method provides quite accurate results in spite of the erroneous observations and a very low sampling rate of 20 samples per second. The maximum value of the random error added usually exceeded 5% of the actual output readings. It is to be noted that when the observations are not erroneous, i.e. when random corruptions are not introduced in the readings of $c_1(t)$ and $c_2(t)$, the accuracy of estimation is precisely 100%. Few, if any, schemes for system identification can boast of a perfect 100% accuracy under ideal conditions.

The process of estimation can actually be implemented by a simple computer program. Of course, the same method can easily be employed to estimate the parameters of an integer order process model as well.

One minor disadvantage of our scheme is that it requires some computational power. But when ultra-high accuracy is needed and the input data is unreliably recorded at a very low sampling rate, this scheme offers a promising solution.